\documentclass[11pt]{article}
\usepackage[inline]{enumitem}
\usepackage[most]{tcolorbox}
\usepackage{amsfonts}
\usepackage{amsmath}
\usepackage{amssymb}
\usepackage{authblk}
\usepackage{array}
\usepackage{bigdelim}
\usepackage{caption}
\usepackage{caption}
\usepackage{centernot}
\usepackage{color}
\usepackage{comment}
\usepackage{dcolumn}
\usepackage{datetime}
\usepackage{diagbox}
\usepackage{eurosym}
\usepackage{footmisc}
\usepackage{float}
\usepackage{geometry}
\usepackage{graphicx}
\usepackage{makecell}
\usepackage{multicol}
\usepackage{multirow}
\usepackage{natbib}
\usepackage{pdflscape}
\usepackage{rotating}
\usepackage{sectsty}
\usepackage{setspace}
\usepackage{standalone}
\usepackage{subfigure}
\usepackage{tikz}
\usepackage{titling}
\usepackage{ulem}
\usepackage{xcolor}
\usepackage[hidelinks]{hyperref}
\hypersetup{unicode = true}
\newcolumntype{d}[1]{D{.}{.}{#1}}
\newcommand{\mc}[1]{\multicolumn{1}{c}{#1}} 

\newtcolorbox{highlighted}{colback=yellow,coltext=red,breakable}

\newcommand{\refp}[1]{(\ref{#1})}

\title{
  Judicial Favoritism of Politicians: \\ Evidence from Small Claims Court\thanks{
    We are grateful to the insightful feedback from Brigitte Seim, Scott Desposato, George Avelino, Ciro Biderman, Ashu Handa, Tricia Sullivan, Doug MacKay, and seminar participants at CEPESP-FGV, the Paran\'a State Bar Association, and the 2019 Meeting of the Canadian Law and Economics Association. We acknowledge financial support from the Institute for Humane Studies (IHS) and the UNC-Chapel Hill Department of Public Policy. All remaining errors are our own.
  }
}
\date{\monthname[\month], \the\year}
\author{
  Andre Assumpcao\thanks{
    Department of Public Policy, The University of North Carolina at Chapel Hill; \href{mailto:aassumpcao@unc.edu}{\textcolor{blue}{aassumpcao@unc.edu}}
  } 
  \and
  Julio Trecenti\thanks{
    Brazilian Association of Jurimetrics; \href{mailto:jtrecenti@abj.org.br}{\textcolor{blue}{jtrecenti@abj.org.br}}
  } \textsuperscript{,}\thanks{
    Institute of Mathematics and Statistics, University of S\~ao Paulo.
  }
}

\begin{document}
\begin{titlepage}

\maketitle

\begin{abstract}
\noindent Multiple studies have documented racial, gender, political ideology, or ethnical biases in comparative judicial systems. Supplementing this literature, we investigate whether judges rule cases differently when one of the litigants is a politician. We suggest a theory of power collusion, according to which judges might use rulings to buy cooperation or threaten members of the other branches of government. We test this theory using a sample of small claims cases in the state of S\~ao Paulo, Brazil, where no collusion should exist. The results show a negative bias of 3.7 percentage points against litigant politicians, indicating that judges punish, rather than favor, politicians in court. This punishment in low-salience cases serves as a warning sign for politicians not to cross the judiciary when exercising checks and balances, suggesting yet another barrier to judicial independence in development settings.\\



\end{abstract}

\setcounter{page}{0}
\thispagestyle{empty}

\end{titlepage}

\clearpage

\section{Introduction} \label{sec:introduction_paper2}
Suppose a case involving a politician is brought before a fair judge. Assume further that lawyers are equally skilled and unable to influence the judge one way or another. Under these simplifying conditions, we can reasonably expect a court decision based on case merits and blind to the political power of litigants. But is this a realistic depiction of judicial decision-making? Besides judges' individual biases, there could exist institutional constraints to a fair trial. In some cases, the judiciary is not an independent branch of government. In others, \emph{de jure} independence is not followed by \emph{de facto} independence. Under separation of powers, the executive and the legislative also have some control over the judiciary, and thus judges could strategically adjust their decisions in exchange for benefits controlled by politicians. In developing countries, where institutional constraints are stronger, the deviations in justice might even be more pronounced. Surprisingly, however, there are not many studies measuring judicial favoritism in cases involving politicians in development settings. This paper fills in the gap by investigating judicial independence when politicians appear before judges in the Brazilian judiciary system.

In recent years, there has been growing interest in the relationship between judicial decision-making and politics in developing countries. Departing from cross-country analyses of judicial independence, recent studies examine the interactions between judges and politicians at the individual level. \citet{Sanchez-MartinezDismantlingInstitutionsCourt2017} is a good example of this. In the study, the author investigates whether defendant employers in Venezuela are more likely to see a favorable outcome if they are affiliated with Venezuela's ruling socialist party. He finds that employers sharing party affiliation with judges are 20 percent more successful at trial.
In \citet{LambaisJudicialSubversionEvidence2018}, the authors identify a 50 percentage point advantage in the win rate at trial for elected versus non-elected candidates when both are defendants in corruption cases in Brazil. These studies provide evidence against judicial independence under different institutional settings. This project supplements these findings by investigating whether favoritism persists in cases where judges and politicians have less at stake (e.g., small claims cases), testing a theory of power collusion across branches of government. We suggest that the judiciary might favor politicians in court in exchange for (or in anticipation of) future benefits controlled by other branches of government.

Besides this theoretical component, this paper makes a series of data analysis contributions to the literature in law and politics. In order to measure the effect of politician bias, we build a unique dataset of small claims decisions from the S\~ao Paulo State Court System in Brazil for all mayor and city council candidates since 2008. We then apply the methodology in \citet{AbramsJudgesVaryTheir2012} to evaluate whether politicians experience any favoritism in their cases. To further investigate the relationship between litigant politicians and judges, we employ a regression discontinuity (RD) strategy comparing decisions across mayor candidates who have barely won and lost a municipal election in the state. Our goal is to verify whether office-holding causes judges to rule in favor of mayors. Finally, we implement various machine learning regression algorithms to predict court outcomes and identify the most critical factors driving bias. To our knowledge, this is the first paper proposing such comprehensive analysis of judicial favoritism of politicians in developing countries.

We find that state court judges rule relatively and significantly less in favor of politicians in small claims cases than not. Litigant politicians experience a 3.7 percentage point smaller probability of winning. Contrary to our expectation, this result suggests that judges use court decisions as a power move to warn politicians and force their cooperation. In the regression discontinuity analysis, we find no difference in rulings for mayor candidates who barely lose or win a municipal election. We identify a positive correlation between vote share and favorable court outcomes, but as we narrow in on smaller vote margins, which suggest more competitive elections, the effect of holding office becomes statistically insignificant at the 5 percent level. Though this result might seem surprising, we believe it is a direct consequence of the structure of local governments in Brazil. Since municipal governments are responsible for implementing high-salience policy, such as health and primary education, mayor candidates enjoy high visibility and might see favorable outcomes even if they do not take up office. In the court outcome prediction exercise, we lastly find that claim amount, judge tenure, pay, and defendant politicians are among the most important factors in judicial decision-making, consistent with indicative evidence from previous studies and with the collusion argument developed here.

These results supplement the empirical legal studies literature in multiple ways. First, we document judicial bias where there should be none. Small claims cases have limited political consequences, so judges have lower incentives to punish politicians. Having documented bias in low salience cases, however, we can conclude that deviations of justice in the state court of S\~ao Paulo are widespread. As legal cases increase in importance, judges and politicians would respond accordingly, and we would expect less independence. Since elections and state court systems are relatively uniform in Brazil, we believe this would be the norm for the entire judiciary system.
By presenting evidence of critical factors driving judicial decisions, we also provide a benchmark for future analysis of politician bias. The amount claimed in court, judge tenure, judge pay, defendant politicians, and politician vote share are the primary factors for judicial decision-making and should be the subject of further scientific investigation. These factors are also first-order issues to be brought to the attention of judges as potential sources of bias, and should significantly help improve state court systems going forward.

The remainder of this paper is as follows. Section \ref{sec:theory_paper2} discusses the theoretical framework behind judicial favoritism and previous evidence; section \ref{sec:background_paper2} presents the institutional design of the court system in Brazil. Section \ref{sec:data_paper2} summarizes the dataset used for analysis; section \ref{sec:methods_paper2} discusses the analytical strategy and presents results. We conclude in section \ref{sec:conclusion_paper2}.

\section{Motivating Judicial Favoritism of Politicians} \label{sec:theory_paper2}

Assume there are three representative public agents, one for each branch of government: the executive, the legislative, and the judiciary. Though these agents are independent, they interact with one another over time. The executive agent serves on one or two four-year mandates (pending reelection at the end of the first term). They control the majority of government budgets, and have the discretion to set wages and allocate resources to the other branches. The legislative agent serves on a two or four-year period,\footnote{Senators serve for eight years but this does not change the theoretical predictions in this section.} but their mandate is renewable as many times as they are reelected; they have no term limit. They are responsible for passing law and determining budget levels but not its composition. In other words, they approve the amount of money available for other branches of government but do not have a say on how to spend the money. The judiciary agent serves on life-long mandates and yields power in restrictive but steady ways. They have limited control over resources as they only oversee budgets in the courts at which they serve, but resolve disputes between the other two branches of government and other economic agents (individuals, companies, etc). In this simplified model, the judiciary interacts with the executive and legislative by settling their disputes.

We are interested in the behavior of the judiciary with respect to other branches of government. Power collusion could exist if the judiciary were using sentences to please the executive and legislative as a means of buying out those who make the calls on judiciary resources. Under this hypothesis, the representative judiciary agent derives utility in each period $t$ according to equation \refp{eq:u_jud}, which describes the benefit $f$ as a function of $k$ observable characteristics $\sum_{x = 1}^{k}x_{k}$, such as their time in post, their wages, their working conditions, and unobservable characteristics $\varepsilon$, such as reputation and their happiness in serving justice; costs $c$ are a function of $m_{t}$ working conditions, executive and legislative utilities $u_{e}^{t}$ and $u_{l}^{t}$; $\delta_{t}$ are exogenous, stochastic shocks that impact judicial work. These per-period utilities are computed in perpetuity (discount factor $r$) in accordance with a judge's mandate:
\begin{equation} \label{eq:u_jud}
  u_{\text{j}}^{t} = \frac{1}{r} \times \sum_{t = 1}^{t} \Bigg[ f\Big(\sum_{x = 1}^{k}x_{k}, \ \varepsilon_{t} \Big) - c\Big(\sum_{m = 1}^{k} m_{t}, \ u_{e}^{t}(p), \ u_{l}^{t}(p), \ \delta_{t}\Big) \Bigg]
\end{equation}

Since the executive and the legislative agents have primary responsibility over government budget, we can expect the judiciary to strategically maximize future net benefit by pleasing other agents. If the executive and the legislative agents are less likely to lose court cases, their utility increases. This incentive, however, is not uniformly distributed across all types of judicial cases. Because electoral, political, and even legal costs are higher in high-salience cases, such as corruption, politicians expect more favoritism in such cases than in low-salience cases. This makes small claims cases, which are one example of relatively less important court cases, an ideal subject for judicial independence research. Any sign of favoritism in these cases is likely carried over to high profile cases, indicating a widespread deviation of justice. In addition, small claims cases are more homogeneous, and suffer less from unobservable factors which could drive court outcomes. In the majority of cases, attorneys have limited ability to influence a case outcome, defendants pay a small monetary compensation if they lose in court, and the duration of cases is shorter than in other courts. In other words, these cases are less likely to suffer from unobservable heterogeneity, such that we can more accurately test whether $\partial u_{j}^{t} / \partial u_{e, \, l}^{t} = 0$. 

\subsection{Previous Evidence on Judicial Bias} \label{subsec:evidence_paper2}

A key principle of liberal democracies is separation of powers. It creates mechanisms of checks and balances aiming at preserving individual liberties and avoiding abuse of government power against its citizens \citep{PerssonSeparationPowersPolitical1997,LaPortaJudicialChecksBalances2004}. An important issue in the literature is whether courts are independent, and if an independent judiciary improves social, political, and economic outcomes. The ability to turn rights and principles into social welfare is a fundamental element for the support of political regimes. In fact, much of the backlash against liberal democracies in recent years is the result of policies that fail to address increasing disparities in social conditions across the world. Judicial independence, however, is not driving the disbelief in democracy: there is documentation of positive impact on growth \citep{VoigtEconomicGrowthJudicial2015}; on the enforcement of contracts and the protection of private property \citep{AcemogluUnbundlingInstitutions2005}; on the control of the bureaucracy \citep{McCubbinsJudiciaryRoleLaw2009,HanssenIndependentCourtsAdministrative2000}; on alternation of power \citep{RamseyerPuzzlingDependenceCourts1994}; and even on the duration of public policy \citep{HanssenTherePoliticallyOptimal2004}.

To fully capture the relationship between judicial independence and social outcomes, a few recent studies have shifted to a disaggregated, individual-level analysis of judicial bias. These studies focus on ethnicity, race, gender, or religion bias \citep[and many others]{AbramsJudgesVaryTheir2012,ShayoJudicialIngroupBias2011,ArnoldRacialBiasBail2018}, but there are only a few tests on the political status of litigants: \citet{HelmkeLogicStrategicDefection2002} documents strategic defection of judges in their rulings in Argentina when governments are weak; \citet{Sanchez-MartinezDismantlingInstitutionsCourt2017} looks at favoritism of defendant government agencies in Venezuela when the agency head is a member of the ruling socialist party; \citet{LambaisJudicialSubversionEvidence2018} examine bias in corruption cases against politicians in Brazil; \citet{Poblete-CazenaveCrimePunishmentPoliticians2019} investigates whether office-holding impacts court outcomes using a sample of state legislators in India. In line with these studies, we also analyze political favoritism at the individual level, using litigation data from S\~ao Paulo, Brazil. However, we extend these studies by testing for bias in cases in which politicians are plaintiffs -- in addition to cases in which they are defendants. The limitation of previous individual-level studies is only investigating the behavior of judges in cases filed \emph{against} but not \emph{by} politicians. This distinction is relevant if there is any selective litigation \citep{PriestSelectionDisputesLitigation1984,WaldfogelSelectionHypothesisRelationship1995}, in which case we can reasonably expect heterogeneous favoritism effects conditional on politicians being  plaintiffs or defendants. Secondly, we look at low-salience cases, where no favoritism is expected. The reasoning is straightforward: since stakes are lower in less relevant cases, we would anticipate more judicial independence. Any bias in these cases would carry over to more salient cases (e.g., corruption) because their higher cost makes politicians more wary and likely to further pressure the judiciary to rule in their favor. Therefore, the evidence here advances our knowledge on bias in different types of litigation.

\section{Court Systems in Brazil} \label{sec:background_paper2}

Brazil's judiciary system is divided into general and limited jurisdiction courts. Federal and State Courts form the general system and Electoral, Military, and Labor Courts form the limited jurisdiction system. There are up to three instances of judicial review in either system, and the court of last resort is the Federal Supreme Court (STF). It takes up cases under its jurisdiction as set out by the Brazilian Constitution, cases in which there are conflicting norms or jurisprudence issued by lower courts, and cases where there is a direct violation of constitutional norms. To limit the sources of heterogeneity, this paper focuses on cases heard in state court systems. In particular, we focus on the state of S\~ao Paulo, the most economically and politically important state in the country.

There are 319 judicial districts in the state, and each district has one or more courthouses. These courthouses host at least one judge with either a broad mandate, meaning that they can rule on any issue within the state court system's attribution, or a narrow mandate, which means they only oversee certain types of cases within the system, e.g., commercial or family law. Within the state system, there are specialized small claims courts called \emph{Juizados Especiais Cíveis} (Special Civil Tribunals, in free translation, and SCTs henceforth). SCTs replaced the primary small claims courts across Brazil upon the passage of the most recent Brazilian Constitution in 1988.\footnote{More evidence of this in \citet{LichandAccessJusticeEntrepreneurship2014}.} Their goal is to simplify and increase access to justice across states by removing many procedural requirements present in other litigation systems. SCTs are the primary judicial body for low-complexity cases, defined as cases in which claims do not exceed 40x the minimum wage.\footnote{There are no state minimum wages in Brazil, so this is the federal, nationwide minimum wage of R\$ 1,040 per month beginning in 2020. This makes the maximum claim amount R\$ 41,600, or $\sim$\$9,900 in current dollars using the 2019 end-of-year exchange rate.} The legal issues they take up range from lease breach, consumer rights, and debt to executions and torts. There is no need for hiring an attorney if claims are under 20 times the minimum wage. SCTs are only open to individual or small company plaintiffs.

An example helps illustrate a typical SCT case. Suppose your mobile phone service provider has been overbilling you for international phone calls that were never made. You, unfortunately, could not resolve this issue with the company's customer service and now want to take legal action and receive financial compensation for the wrongful charges. You walk up to an SCT office, speak to a courthouse clerk, and file your claim along with any supporting documentation. The clerk then provides a court date for a conciliation hearing. At the hearing, you and the phone company will try to reach an agreement; if that fails, the judge sets trial for either later that same day or in the following days. At trial, the judge issues a sentence which can be appealed within ten days; on appeal, a three-judge panel then issues the final ruling. This entire process might take less than three months, representing a substantial improvement when compared to cases in the regular judicial process at state courts, which take up to 38 months on average to conclude \citep{CNJJusticaEmNumeros2018}.

SCT's structure dramatically reduces the number of dimensions driving judicial decisions. According to the S\~ao Paulo State Court website, there are less than 15 types of cases that can be brought before SCTs. It is then easier for judges and lawyers, when hired by the litigants, to specialize and reduce any skill discrepancy that could substantially drive a case outcome. The settlement rate is about 4 percent higher in  SCTs \citep{CNJJusticaEmNumeros2018} compared to all other court systems. The sentence is also standard across cases: the losing side will pay the claim amount to the winning side, which is capped at 40x the minimum wage. The standard and relative low-salience punishment removes additional sources of heterogeneity from high-profile cases, such as corruption cases in \citet{LambaisJudicialSubversionEvidence2018} or violent crimes in \citet{LimJudgePoliticianPress2015}. Small claims cases do not attract as much media attention as corruption or violent crime, for instance. In fact, the use of these cases is an approach first introduced by \citet{ShayoJudicialIngroupBias2011} that takes advantage of the relative homogeneity of small claims in Israel to isolate the effect of ethnicity and religion on court outcomes. Lastly, judges have no control over which SCT cases they take. In single-judge courthouses, all cases are presented before the same judge; in multiple-judge courthouses, the cases are randomly distributed to judges serving on the same SCT. These distribution rules reduce the sources of external variability of outcomes and prevent cases from being differently assigned to systematically more lenient (or harsher) judges in the state system.\footnote{Yet, for robustness purposes, we replicate the process in \citet{AbramsJudgesVaryTheir2012} producing random distributions of court outcomes to serve as a check on the quality of the case assignment system implemented by the state of S\~ao Paulo.}

\section{Data} \label{sec:data_paper2}

We construct a case-level dataset (5,224 observations) with information from two sources. First, we collect SCT case and judge information from the S\~ao Paulo State Court (TJ-SP). The court publishes all judicial decisions on their website, and the information available is the case duration, type (breach of contract, debt execution, and others above), the court where it was filed, ruling judge, the claim amount, litigants and their lawyers (if hired), and sentencing documents. Second, we use the information on candidates running for municipal office in the state of S\~ao Paulo in 2008, 2012, and 2016 from the Brazilian Electoral Court (TSE). TSE has jurisdiction over the entire electoral process in Brazil, from registering candidates, ruling over breach of electoral law, and overseeing the voting process on election day, to counting votes and authorizing that elected politicians take up office. It collects individual-level data on politicians and publishes everything online. We use TSE electoral results, candidate information, and electoral district data for every elected candidate in the state of S\~ao Paulo in the municipal elections of 2008, 2012, and 2016. Table \ref{tab:sumstats_paper2} contains the descriptive statistics of this sample of candidates in the state.

Though the targeted case duration is three months, we can see that cases will last on average 361 days to conclude (12 months), and the average amount claimed by plaintiffs is R\$11,750 ($\sim$US\$2,790). In 64.5 percent of the cases, judges rule in favor of politicians. Sixty-one percent of judges are male and have held their position for over ten years. Their income is R\$35,152 per month, or US\$8,370, on average. The average age of candidates involved in SCT cases in the state is 44 years old, 90.1 percent are male, and 24.3 percent have previous political experience, measured by an indicator variable for candidates who have been reelected or have declared their occupation to the Electoral Court as politician of any kind (city councilor, mayor, governor, member of Congress, senator, president). Forty percent of the politicians in our sample were holding office at the time these cases were heard by the state court. On average, they spent R\$18,212 (US\$4,336) on the campaign trail. We have also collected categorical variables for educational attainment and marital status for all elected officials, but they are omitted from the table. The most frequent educational level and marital status are a four-year college degree or equivalent (41.2 percent) and married (70.3 percent), respectively.

\section{Empirical Analysis}\label{sec:methods_paper2}

We analyze judicial favoritism in five steps. We start off by confirm random assignment of cases. Next, we test whether the distribution of rulings in favor of politicians resembles a random distribution of court rulings. Any deviation would indicate judges do vary how they try cases having politicians as litigants. Third, we estimate the causal effect of holding office using a regression discontinuity design. This result is supplemental to judicial favoritism because it allows us to check heterogeneous effects for those who run for office and those who run and take up office. Step four is checking  bias results against a sample of similar small claims cases, decided by the same judges around the same time, but whose litigants are not politicians. Our goal is to provide a counterfactual estimate of judicial outcomes in the absence of litigant politicians. Finally, we use four machine learning algorithms to predict court outcomes and recover the most critical factors driving judges' decisions in SCT cases.

\subsection{Are Cases Assigned at Random?}\label{subsec:random_paper2}

To reasonably compare court outcomes across judges, our first concern is establishing that judges cannot select into cases they try. The design of the Brazilian state court system in general and SCTs in particular is the first guarantee in support of random assignment of cases. First, plaintiffs file small claims lawsuits where the wrongful act took place or where they (or the defendant) live. Judges do not choose whether to hear the case. Plaintiffs walk up to the court, speak to clerks, and are instructed on how to proceed. State judges come in only when the case has been opened by clerks and cannot choose which of the open cases to try. Second, in districts where there is more than one judge trying cases, the allocation of lawsuits by judge is random; the distribution of cases across judges happens immediately after the case has been included in the caseload management system. Together, these two legal features indicate limited, if any, control of case allocation across judges.

Nevertheless, we provide empirical support to random case assignment. We follow the strategy in \citet{AbramsJudgesVaryTheir2012}, who analyze racial bias in felony cases sentencing across judges in Cook County, IL. They suggest that random assignment could be tested regressing a case characteristic (e.g., politician's age) on multiple control variables, as below:
\begin{equation} \label{eq:methods1}
  \text{Age}_{ijc} = \alpha + \sum_{k=1}^{k} \beta x_{kijc} + \sum_{n=1}^{n} \Gamma D_{n} + \sum_{c = 1}^{c} \lambda_{c} + \varepsilon_{ijc}
\end{equation}

Where Age$_{ijc}$ is the politician age, $x_{k}$ are $k$ control variables, $D_{n}$ is a matrix of $n$ judge fixed-effects, $\lambda_{c}$ is a matrix of court fixed-effects, and subscripts $i$, $j$, $c$ are indexing case $i$, judge $j$, and court $c$. Under random assignment, the \emph{F}-test on the joint distribution of judge fixed-effects should fail to reject the null (i.e.,~fixed-effects have the same effect on politician's age). In other words, it means that there is no substantial correlation between judge fixed-effects and age that would indicate systematic assignment of cases for politicians of a certain age to any judge. However, this direct \emph{F}-test would likely lead to overrejection of the null as described by \citet{AbramsJudgesVaryTheir2012}. Since the number of judges per SCT court and the number of cases per judge are relatively small (an average of 1.74 judges per court and 10 cases per judge), this \emph{F}-test would not meet the asymptotic properties of the \emph{F}-statistic. It would suffer from finite-sample bias (the judge variability within courts is small), and thus we would be rejecting the null effect of judge fixed-effects because of the data-generating process rather than the true null effect.

The solution to the overrejection problem is the construction of simulated datasets where the assignment of cases is indeed random and the subsequent comparison of statistical moments in the empirical distribution versus the simulated moments. Since the data-generating process is random by construction, the \emph{F}-statistic unequivocally provides evidence of true judge fixed-effects. This exercise is as follows. First, the researcher should group the actual (empirical) sample into the randomization units, which are the clusters within which the distribution of cases is random. In this study, these units are the  SCT courts, and they have one or more judges on the bench. The researcher then recreates the distribution of cases within the unit by randomly drawing, with replacement, cases from the unit and reassigning them across judges. This process creates a simulated (random) sample of cases mirroring the empirical sample. An example helps further illustrate this point. Suppose four judges heard 20 cases (five each) in a given SCT in the state of S\~ao Paulo. Each case has a set of observed characteristics, e.g., plaintiff gender, age, claim type, claim amount, whether the politician was the plaintiff or defendant litigant, and so on. The researcher then creates 20 simulated cases, five per judge, keeping the same proportion as in the original data, where each case characteristic is randomly drawn from the sample of 20 observed cases. Once this process is replicated for all randomization units (SCTs), a simulated dataset of the same size as the empirical dataset has been created. We finally repeat this process to create 1,000 simulated datasets.

\begin{figure}[H]
\centering
\caption[figure1]{Interquartile Range of Candidate Age by Judge}\label{fig:randomassignment1}
\vspace{-10pt}
\includegraphics[scale = .60]{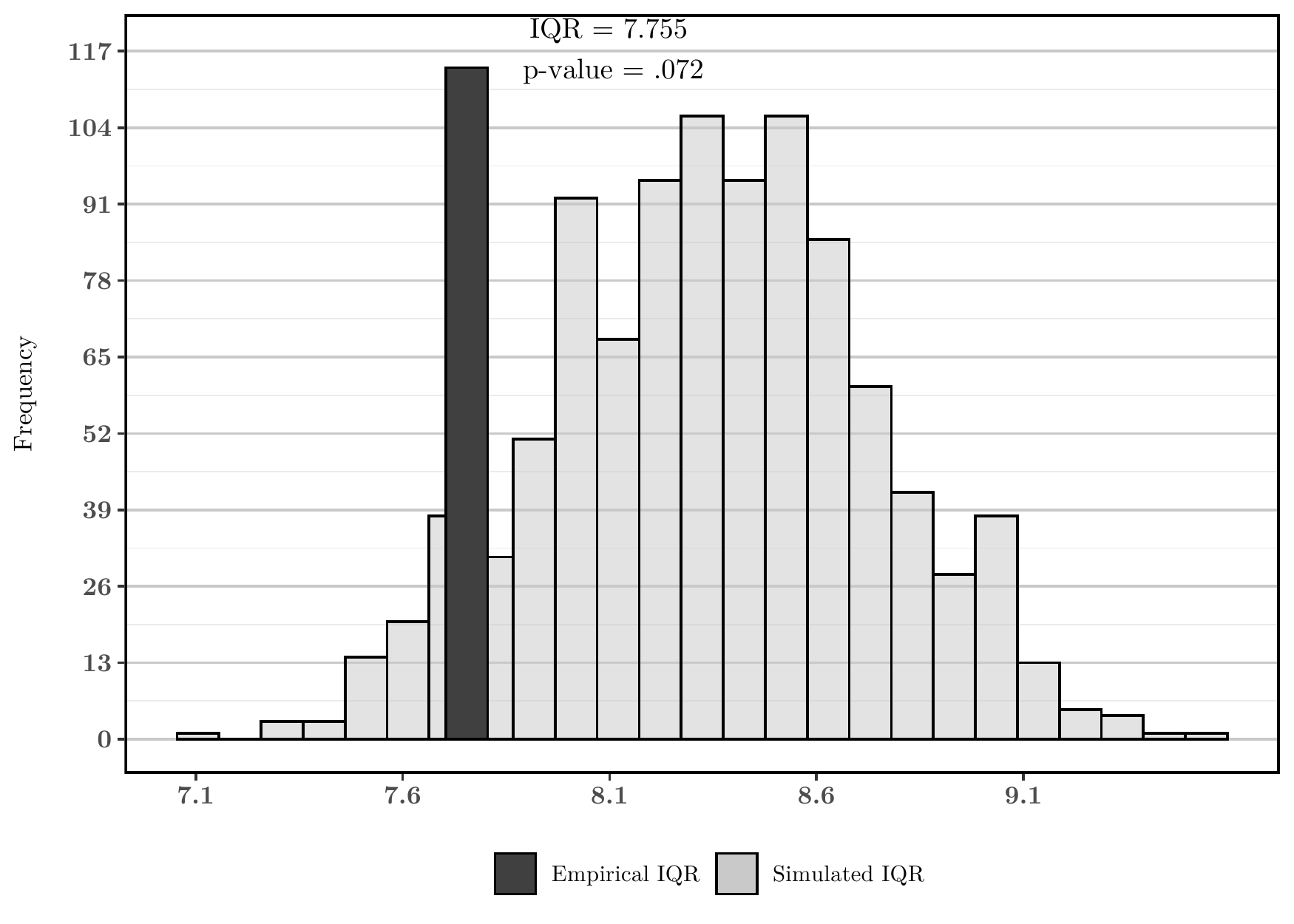}
\end{figure}

Armed with these datasets, we compare the statistical moments in the empirical and simulated datasets without fearing overrejection. Similarly to \citet{AbramsJudgesVaryTheir2012}, we compute the means of politician age for all cases tried by any judge and compute the 25-75 interquartile range (IQR) for the entire distribution of judges. We report the results in figure \ref{fig:randomassignment1}. The empirical IQR for age is 7.75. This value summarizes the difference in politician age for the middle 50 percent of cases across all judges. If this empirical moment is not statistically different from a simulated distribution of politician age IQR, then there is evidence in support of the random allocation of cases -- which is what we expect from the design of SCTs in S\~ao Paulo. In figure \ref{fig:randomassignment1}, the \emph{p}-value for an IQR of 7.75 in the simulated distribution is .072, such that we indeed find support for random assignment at the 5 percent level. Though one could claim that .072 is very close to the .05 threshold, the small number of cases per judge restricts our ability of increasing the variance of case characteristics within each randomization unit. Our average pool for randomizing case characteristics is small (10 cases per judge) compared to \citet{AbramsJudgesVaryTheir2012} (489 cases per judge). Therefore, small increments in cases per judge would substantially increase variance, shifting the empirical realization of IQR to the right, to a higher \emph{p}-value. If anything, the empirical IQR \emph{p}-value remains underestimated, lending support to random assignment of cases in S\~ao Paulo's SCTs.

\subsection{Heterogeneous Convictions Across Judges}\label{subsec:iqr_paper2}

\begin{figure}[H]
\centering
\caption[figure2]{Interquartile Range of Favorable Ruling by Judge}\label{fig:randomassignment2}
\vspace{-10pt}
\includegraphics[scale = .60]{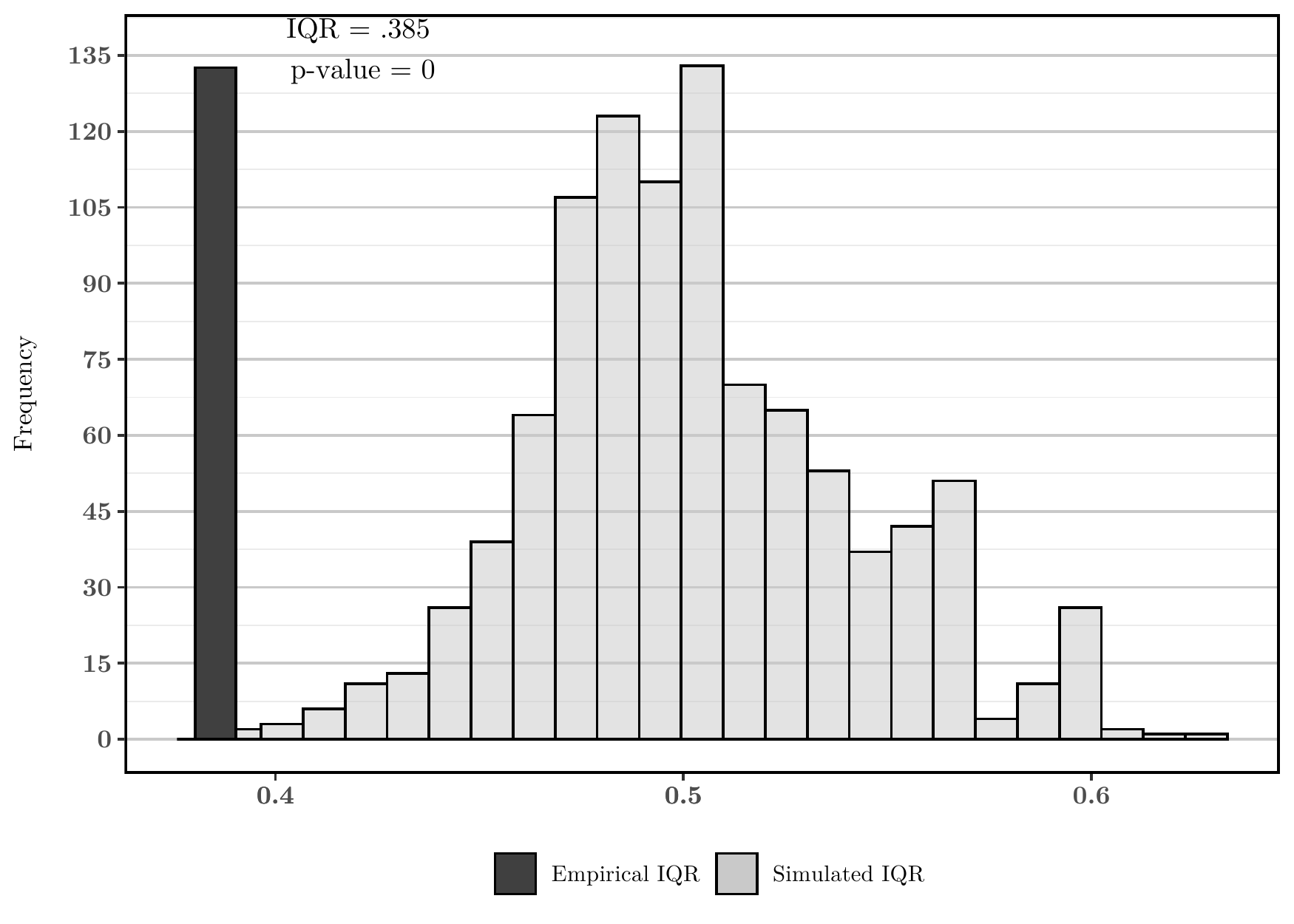}
\end{figure}

To measure judicial favoritism, we carry out a similar process. We produce the same simulated datasets and IQRs, but instead of looking at the distribution of a case covariate, we examine the distribution of the outcome of interest, i.e., whether the case has been ruled in favor of a politician. We present the results in figure \ref{fig:randomassignment2}, where we plot the simulated IQR range across all judges in gray. These judges vary in the harshness with which they rule cases involving politicians. The IQR can thus be interpreted as the difference in politician win probability across the harsh-lenient judge spectrum. In the simulated data, this harshness distance is 50.3 percentage points. It means that if a politician's case were shifted away from a harsh (lenient) judge, their win probability should have increased (decreased) by 50.3 percentage points. We expected this simulated result: a uniform distribution of rulings predicts the 25 percent harshest (most lenient) judges trying in favor (against) of politicians 25 percent of the times. The 50 percent in the middle of the harshness spectrum is the simulated IQR. In this set-up, the distribution of win rates matches the quantiles of the spectrum.

\begin{figure}[H]
\centering
\caption[figure3]{Comparison of Politician Win Probabilities Per Judge}\label{fig:sct_comparison}
\vspace{-10pt}
\includegraphics[scale = .60]{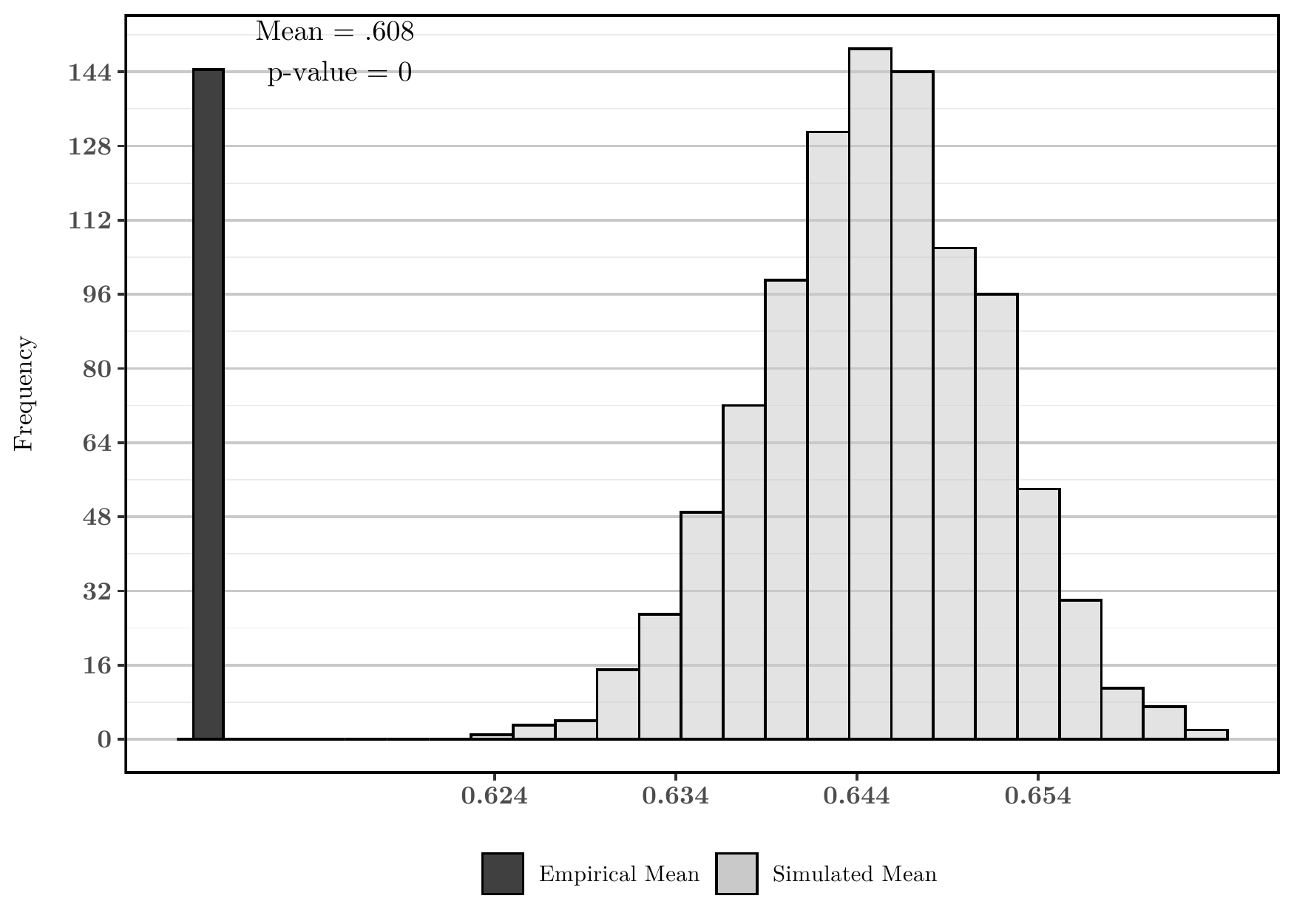}
\end{figure}

The empirical IQR, however, is 38.5 percentage points. This distance between the more strict and more lenient judges is statistically different (\emph{p}-value $= .0001$), and smaller, than the simulated IQR. There is less variation in court outcomes in the empirical data when compared to the simulated data, which means that the distribution of court outcomes for litigant politicians is not random. Judges at the tails of the empirical distribution do not rule as expected. However, we do not know the direction of the court decisions, whether judges' are handing out harsher or more lenient rulings. In other words, it means we still need to answer whether these systematic decisions are in favor or against politicians. To answer this question, we examine the politician win rate, rather than the IQR, across all 514 judges. The results are presented in figure \ref{fig:sct_comparison}, where we run the same simulation exercise and compare the empirical distribution of pro-politician decisions against a simulated distribution of pro-politician decisions. We use the same process as before to reflect judges' heterogeneous harshness preferences, in which we draw outcomes for each judge from the same randomization units (the SCT courts). In figure \ref{fig:sct_comparison}, we plot a empirical win rate of 60.8 percent of all cases against the distribution of simulated win rates (mean $= 64.5$ percent; standard deviation $= .006$). Thus, we document a 3.7 percentage point lower win rate for politicians compared to simulated win rates. This result is evidence of systematic punishment of politicians in SCT court.

Contrary to our expectation, state judges do not favor politicians in anticipation of future benefits. Indeed, they are harder than expected on politicians. A potential explanation for this result is that these unfavorable decisions are judicial demonstrations of power. Judges punish politicians in these low-salience cases as warning signs of future hostility in case politicians pass on their judicial benefits. The judiciary is still using decisions as means of extracting benefits, but rather than using a carrot (favorable decisions), they are using a stick (unfavorable decisions) to keep politician behavior in check. Unfortunately, we cannot test this mechanism further unless we have more on judicial decisions and the involvement of politicians in court cases.

\subsection{Does Holding Office Matter?}\label{subsec:rd_paper2}

In addition to the negative effect documented above, an interesting question is whether judges respond differently to politicians who are holding office at the time of trial. Since office-holding politicians yield more power compared to politicians seeking office because the former are effectively making policy decisions, a plausible hypothesis is that judges would target elected politicians more often. The ideal experiment to answer this question would be to randomly assign office across politicians involved in court cases and compare their outcomes. We would be able to compare how elected officials fare against non-elected officials and pin down the exact bias in favor or against office-holders. For ethical reasons, however, this is an unrealistic experiment, and the regression discontinuity design we adopt in this paper is its best approximation yet.

We start by subsetting our sample to only include mayor candidates. They face each other off in majoritarian elections, and the candidate who reaches 50 percent plus one of the votes is elected to office. This design allows us to compute the vote share for each mayor candidate and compare to the 50 percent threshold: politicians elected to office with narrow margins approximate random sorting to office when compared to politicians who barely lost their election. Vote margin is the running variable for a standard regression discontinuity design where the treatment condition is holding elected office at the time of trial. \citet{LeeRandomizedexperimentsnonrandom2008} inaugurated this research approach, and there are many studies that adopt such strategy to measure social, political, or economic outcomes in Brazil \citep[][and many others]{BrolloTyingYourEnemy2012,DeMagalhaesIncumbencyEffectsComparative2015,BrolloWhatHappensWhen2016}. Formally, we estimate the following equation:
\begin{equation} \label{eq:rddregression}
  \begin{aligned}
    y = \alpha + \gamma_{1}(x-c) + \gamma_{2}(x-c)^{2}+ \rho_{1}(x \geq c) + \rho_{2} (x-c) (x \geq c) + \varepsilon
  \end{aligned}
\end{equation}

\begin{figure}[H]
\centering
\caption[figure4]{The Causal Effect of Election on Court Outcomes}\label{fig:rdplot}
\vspace{-10pt}
\includegraphics[scale = .60]{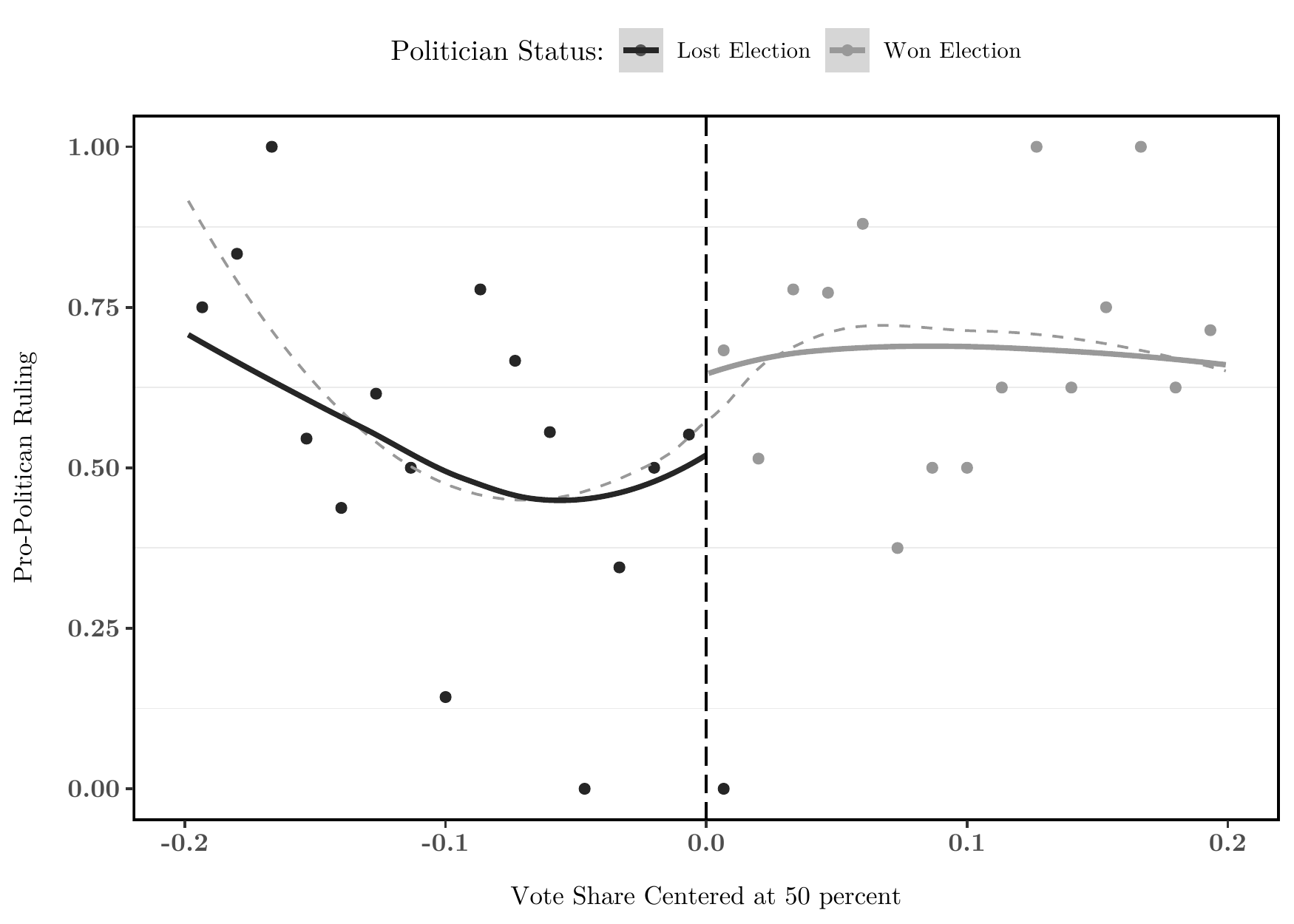}
\end{figure}

Where $y$ is whether the elected politician has received a favorable SCT ruling; $(x - c)$ is the vote share centered at 50 percent, and the squared term captures decreasing returns to scale in vote share; $(x \geq c)$ is the treatment indicator variable, i.e., whether the candidate has been elected; the interaction $(x - c)(x \geq c)$ summarizes differential trends on both sides of the vote share threshold. Using the bandwidth selection model from \citet{CalonicoOptimalDataDrivenRegression2015}, $\rho_{1}$ represents the causal effect of holding office on SCT ruling. Figure \ref{fig:rdplot} plots the results of equation \ref{eq:rddregression} using the optimal bandwidth of 8.4 percentage points. Though there seems to exist a positive correlation between holding office and favorable SCT rulings, the causal effect is not significant at \emph{p}-value $= .05$. Mayor candidates who barely lost or won an election seem to perform equally well in court. We force a discontinuity at vote margin zero to make the causal effect explicit, but the potential causal impact is just masking the non-linear relationship between $x$ (vote margin) and $y$ (court ruling). The dashed line depicts the polynomial regression on the entire sample and serves as evidence of this non-linearity.

\begin{figure}[H]
\centering
\caption[figure5]{Election Point Estimates (and their 95\% CIs) \\ on Court Outcomes}\label{fig:rdbws}
\vspace{-10pt}
\includegraphics[scale = .525]{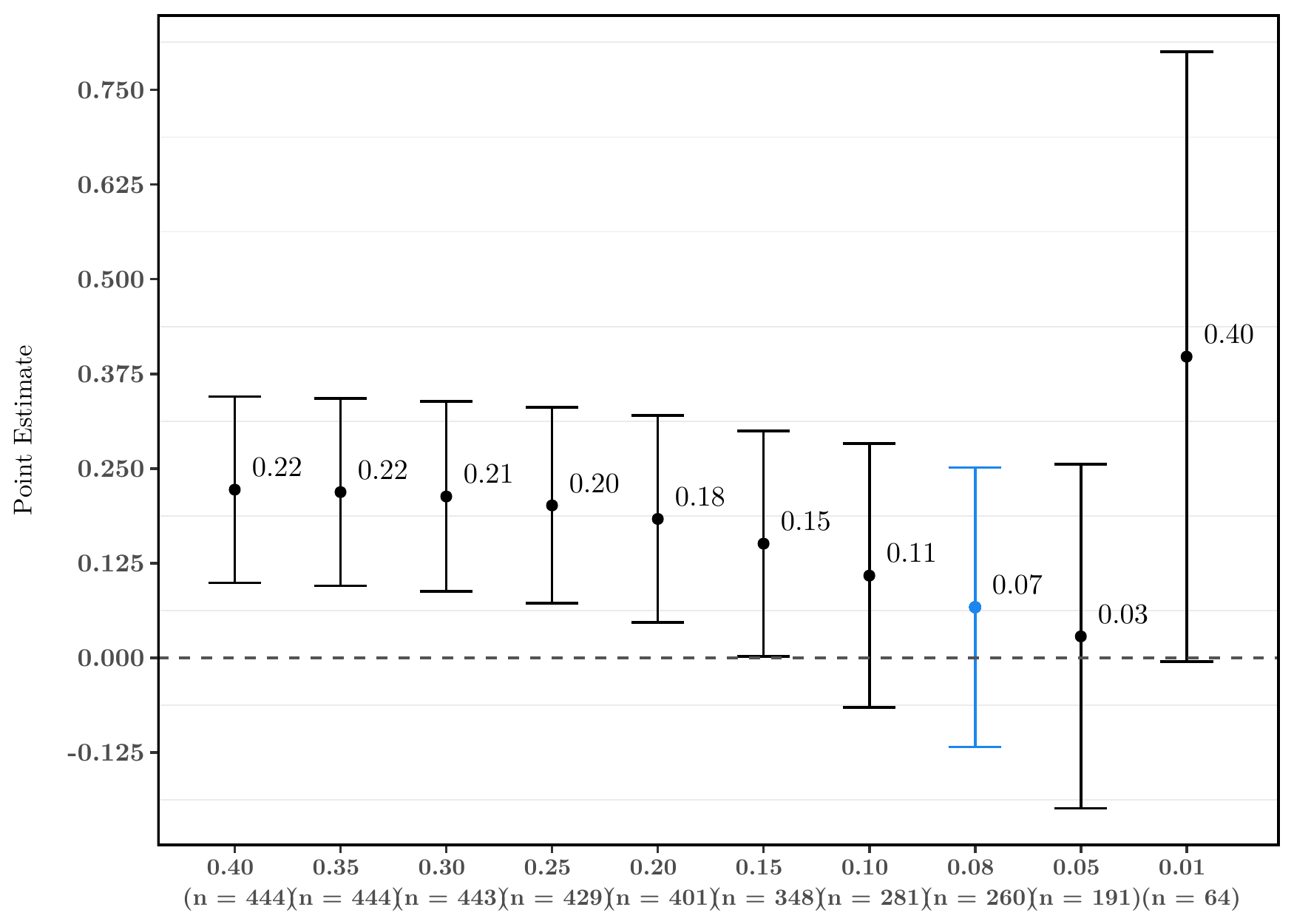}
\end{figure}

Figure \ref{fig:rdbws} presents a more detailed picture of the relationship between politicians and judges in court cases. We re-estimate equation \ref{eq:rddregression} using different bandwidth sizes \citep[as suggested by][and others]{LeeRegressionDiscontinuityDesigns2010}, which are reported on the \emph{x}-axis. The \citet{CalonicoOptimalDataDrivenRegression2015} optimal bandwidth is again 8.4 percentage points (in blue). The \emph{y}-axis displays the point estimates for $\rho_{1}$ in each regression equation. The results confirm a positive and significant relationship (at \emph{p}-value $= .05$) between vote share and favorable court outcomes across the wider bandwidths (40 through 15 percentage point margins). However, since the significant effect is only present in larger bandwidths, and candidates in these vote margins are not likely comparable to each other, we cannot conclude this effect is causal. As we narrow in on smaller margins, the election coefficient becomes insignificant.\footnote{Despite the larger causal effect at the smallest margin (1 percentage point), the small sample size prevents us from making stronger inference claims about the relationship between holding office and seeing a favorable result in court.}

Though not causal, there is evidence of a positive and significant association between vote margin and SCT outcome. We cannot recover the unbiased, causal effect of holding office on court outcomes.\footnote{We also find null causal effects using difference-in-differences as an alternative identification strategy. We use an indicator variable for trial decision issued after previous election (time variable) and another for when politicians were elected to office (treatment variable). Their interaction, i.e.,~the causal effect, is null regardless of the model we estimate (including or excluding covariates and fixed-effects). These models and their results are available upon request.} A straightforward explanation to the null effect is the visibility of mayors in Brazil. Since municipal governments are in charge of health and education policies, mayor candidates are well-known local figures (much more important than counterparts in U.S.~local elections). Candidates run for office multiple times and could be perceived as politically important even when they are not in office. This visibility would explain why office might not matter for court outcomes. Additionally, though the positive correlation here might seem at odds with the negative bias in section \ref{subsec:iqr_paper2}, we note that the sample used for the causal test in this section is only composed of mayor candidates. The original sample contains individuals running for both mayor and city council seats. It seems possible that judges adjust their decision-making conditional on these power differences across politicians, and that city council candidates are punished more harshly. Unfortunately, we cannot perform the simulation exercise on the mayor-only sample due to the small number of observations in this group ($n = 489$). We would run an even higher risk of overrejecting the null than before.

\subsection{Comparing Politician vs.~Non-politician Rulings}\label{subsec:robust_paper2}

An important concern with our results is the extent to which our sample is representative of the universe of decisions handed out by SCT judges in the State of S\~ao Paulo. If there is any selection into certain SCT courts, then we might be incorrectly estimating the true win probability for politicians. For instance, state judges serving at selected courts in our sample might be naturally harsher, or have a lesser opinion of politicians, and thus the outcomes we observe reflect a single, biased draw from the distribution of SCT courts across the state. The results would not be externally valid.

To address these concerns, we recreate the analysis in section \ref{subsec:iqr_paper2} using another sample of small claims cases, decided by the same judges around the same time, but whose litigants are not politicians. Except for the absence of litigant politicians, these cases share all other characteristics with the primary sample. We construct this sample using the unique identification number assigned to all court cases, in all states, as required by law in Brazil. This case ID is a sequence of 20 digits structured as follows: the case filing order (7 digits), a control sequence (2 digits), the filing year (4 digits), the court system identifier (3 digits), and a district courthouse identifier (4 digits).\footnote{A courthouse identifier (\emph{fórum}) is the building where the courts are physically installed. In the state of S\~ao Paulo, the majority of judicial districts have just one \emph{fórum}. Hence, the last four digits almost always also identifies the judicial district. The exceptions to this rule are the biggest judicial districts, such as the cities of S\~ao Paulo and Campinas.} To construct the counterfactual sample, we marginally change the filing digits to recover lawsuits filed immediately before and after every politician case, but which were still decided by the same judge at the same time. In other words, we add and subtract one to each filing sequence, keeping all other digits the same, and use these new case IDs to find lawsuits and their information at the S\~ao Paulo state court website. Table \ref{tab:numerocnj} displays a sample ID from the politician sample: case 3002615 was filed in 2013 at court system 8, state 26, district 0510. In this case, the IDs for constructing the counterfactual sample are $\underline{3002614}$--$95.2013.8.26.0510$ and $\underline{3002616}$--$65.2013.8.26.0510$.\footnote{The control sequence changes but its calculation is public, so we reverse-engineer these two digits for all cases.} Since our primary sample has 5,224 observations, we recover 10,448 potential lawsuits to serve as counterfactual units. After excluding invalid IDs, lawsuits for which there is no information available online, and non-SCT cases tried at the same judicial district as SCT cases, our final counterfactual sample contains 3,233 cases.

Table \ref{tab:sumstas_random} displays descriptive statistics of the counterfactual sample (panel A) and the tests for mean differences across original and counterfactual sample (panel B). To compare across groups, we keep the same variables except for the court outcome measure. Since there are no politicians in the alternative sample, we cannot compare pro-politician rulings. The percentage of cases ruled in favor of claimants or defendants, however, can be calculated and compared across samples. We proceed with the former. Panel B shows significant differences (at \emph{p}-value $= .05$) for case claim amount and the share of claimant win, but no differences in judge characteristics across samples. These case-level variables, however, ignore shared variation across outcomes due to judges trying multiple cases at a judicial district. We thus suggest measuring outcomes at the judge-level, computing average win rates for all judges in the sample. The results of this exercise are presented in table \ref{tab:agg_outcomes}. When win rates are computed across judges, disregarding litigant status, we observe no significant difference between the politician and the non-politician samples in rows one and three (\emph{p}-value $= .538$ for claimant win rate comparison; \emph{p}-value $= .515$ for defendant win rate comparison). Therefore, we are confident that the politician sample is representative of SCT litigation in S\~ao Paulo.

\paragraph{Do Politicians Experience Worse Outcomes as Claimants or Defendants?}

In section \ref{subsec:iqr_paper2}, we document a negative bias against politicians in small claims court decisions. Compared to simulated rulings, we would expect a 3.7 percentage point increase in their win probability if there were no judge fixed-effects. Using the alternative sample of non-politician litigants, we can investigate this effect further. In particular, we want to know if the negative bias has different magnitudes if politicians are claimants or defendants. Previous evidence in judicial favoritism of politicians only focused on cases of defendant politicians.

We do this by comparing win rates for claimant and defendants across the two samples. First, we benchmark win rates by disregarding the status of litigants. Then, we calculate win rates for litigants in the main sample conditioning means on their politician status. We present the evidence in rows two and four of table \ref{tab:agg_outcomes}. There are no differences between claimant politicians and any of the other claimants (\emph{p}-value $= .169$). Their win rate of 72.9 percent is not significantly different than the 75 percent for other claimants. However, in the last row, we document defendant politicians experiencing more losses than other defendants. The win rate for defendant politicians is 21.6 percent of all their cases, while defendants in the alternative sample win 24.9 percent of the times. The difference is significant at \emph{p}-value $= .10$. The worse performance of 3.3 percentage points accounts for almost all magnitude of the negative bias (3.7 percentage points) in section \ref{subsec:robust_paper2}. This evidence is consistent with other studies. \citet{LambaisJudicialSubversionEvidence2018} document a 50 percent decrease in win probability when defendant politicians lose elections. Their effect size is particularly large, but differences in research design and sample composition explain the smaller effect here. In \citet{LambaisJudicialSubversionEvidence2018}, candidates are being prosecuted for corruption, which is a more salient issue for politicians than the small claims in this study. Second, \citet{LambaisJudicialSubversionEvidence2018} only measure court outcomes for competitive mayors, those who closely lost or won elections, while we examine court decisions for all candidates for mayor and city council in the state of S\~ao Paulo. Our sample contains less skilled politicians, who are less likely to influence court outcomes.

\subsection{Predicting Drivers of Court Outcomes}\label{subsec:predictions_paper2}

In the previous sections, we provided evidence against judicial independence in small claims cases involving politicians in Brazil. Politicians face a 3.7 percentage point lower win rate at trial, and the majority of the effect (3.3 percentage points) comes from cases in which politicians are defendants in SCT court. We want to go beyond court outcomes, however, and examine the drivers behind judicial decisions in cases involving politicians. Our goal is to identify whether certain characteristics are better predictors of court rulings, thus informing the sources of the negative bias against politicians in our analysis.

To this end, we deploy four machine learning (ML) regression algorithms and compare how well they predict court outcomes. Predictive models have gained ground in the social sciences \citep{MullainathanMachineLearningApplied2017,AtheypredictionUsingbig2017,AtheyImpactMachineLearning2019}, and we are interested in their ability to point to individual features (variables) that contribute the most for the prediction of outcomes. Following standard practice in ML applications, we split the original sample of politician cases into train and test datasets, using a ratio of 80-20 for the number of observations in each group. The former portion of the data is used to train the algorithms, within which we apply five-fold cross-validation. We feed the algorithm the features (covariates) and the labels (outcome) and ask it to predict the labels based on the features. Once these algorithms have been trained, we repeat the process on the latter portion of the data, also known as \emph{hold-out}, so that we can compare prediction performance across models. Each model will use a different prediction technique, but they will all produce an accuracy score, which is each model's percentage of correct predictions over the entire sample. We also use Cohen's Kappa scores \citep{LandisMeasurementObserverAgreement1977}, which indicate the increment in the rate of agreement between predicted and observed labels on top of random predictors.

Table \ref{tab:prediction} displays the performance measures for each model used in the analysis. The four models are (1) Logistic regression with Lasso regularization \citep{TibshiraniRegressionShrinkageSelection1996}; (2) Random Forest \citep{BreimanRandomForests2001}; (3) Gradient Boosting \citep{FriedmanGreedyFunctionApproximation2001}; and (4) Deep Neural Networks using dense layers \citep{GoodfellowDeepLearning2016}. The best performing model is  Random Forest. It predicts the correct court outcomes in 77 percent of the cases in the hold-out data and performs 19.2 percent better than a random classifier. What we are interest in, however, is going beyond predictions and identifying the underlying factors driving judicial decisions. A key feature of Random Forest regression is that we can verify the most important variables for outcome prediction. Since the model's outcomes are decisions made by state judges in S\~ao Paulo, we can thus identify the factors behind judicial decisions as the most important variables in the Random Forest regression. The most popular variable importance measure is the Mean Decrease in Gini.

\begin{figure}[H]
\centering
\caption[figure5]{Random Forest Variable Importance}\label{fig:rfvarimp}
\vspace{-10pt}
\includegraphics[scale = .60]{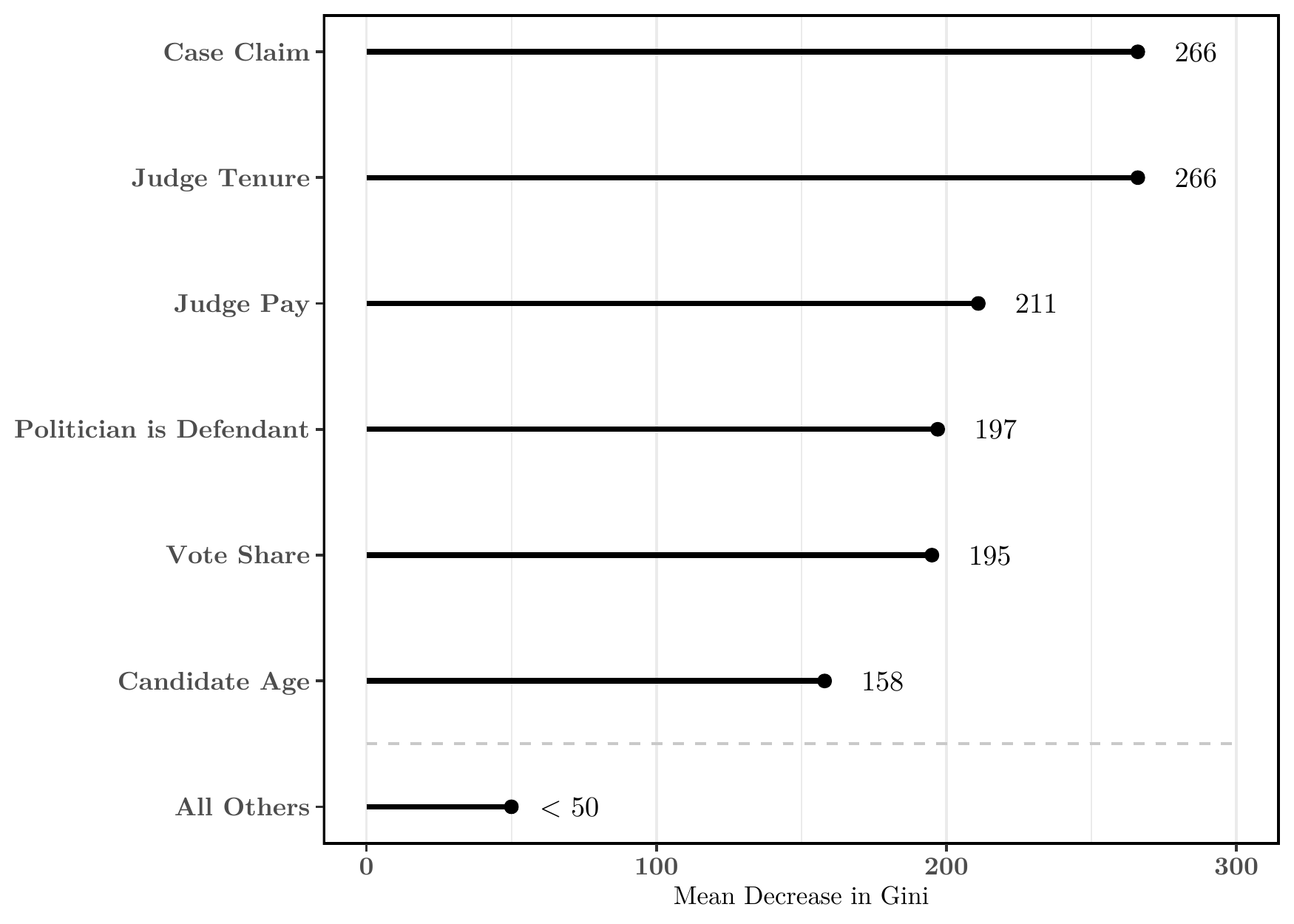}
\end{figure}

The Gini Coefficient of a Random Forest model measures the contribution of each variable to the homogeneity of trees' nodes and leaves. The more homogeneous a tree is, the higher its Gini Coefficient and the model's accuracy (prediction). The Mean Decrease in Gini is just the average of Gini Coefficients for all decision trees used to build the model, and higher values also indicate higher variable importance. Figure \ref{fig:rfvarimp} shows the variable importance plot based on Mean Decrease in Gini. The factors on the \emph{y}-axis are ranked by their score. The most critical factors are amount claimed in court and judge tenure, followed by judge pay, the status of politicians as defendants, candidates' vote shares, and age. We are particularly interested in the political variables; unsurprisingly, judges pay close attention to cases involving defendant politicians, for whom we document the worst performance in court. We also see that judges respond to candidate's vote share. This result signals that decisions in politician cases take into account the relative power of candidates as measured by their vote count. This evidence is consistent with the model in section \ref{sec:theory_paper2}, where we posited that politicians and judges develop a mutually beneficial relationship; i.e., they collude with one another when their utility of cooperation is higher than their utility of competition. For instance, ruling in favor of a politician in an SCT case would create a positive attitude towards increasing resources available to the judicial branch. Contrary to our expectation, however, the mechanism behind cooperation is not favorable decisions; in fact, state judges rule against politicians in low-salience small claims cases, signalling power, to buy cooperation from local politicians.

\begin{figure}[H]
\centering
\caption[figure6]{PDPs for the four most critical predictors}\label{fig:pdp}
\vspace{-10pt}
\includegraphics[scale = .65]{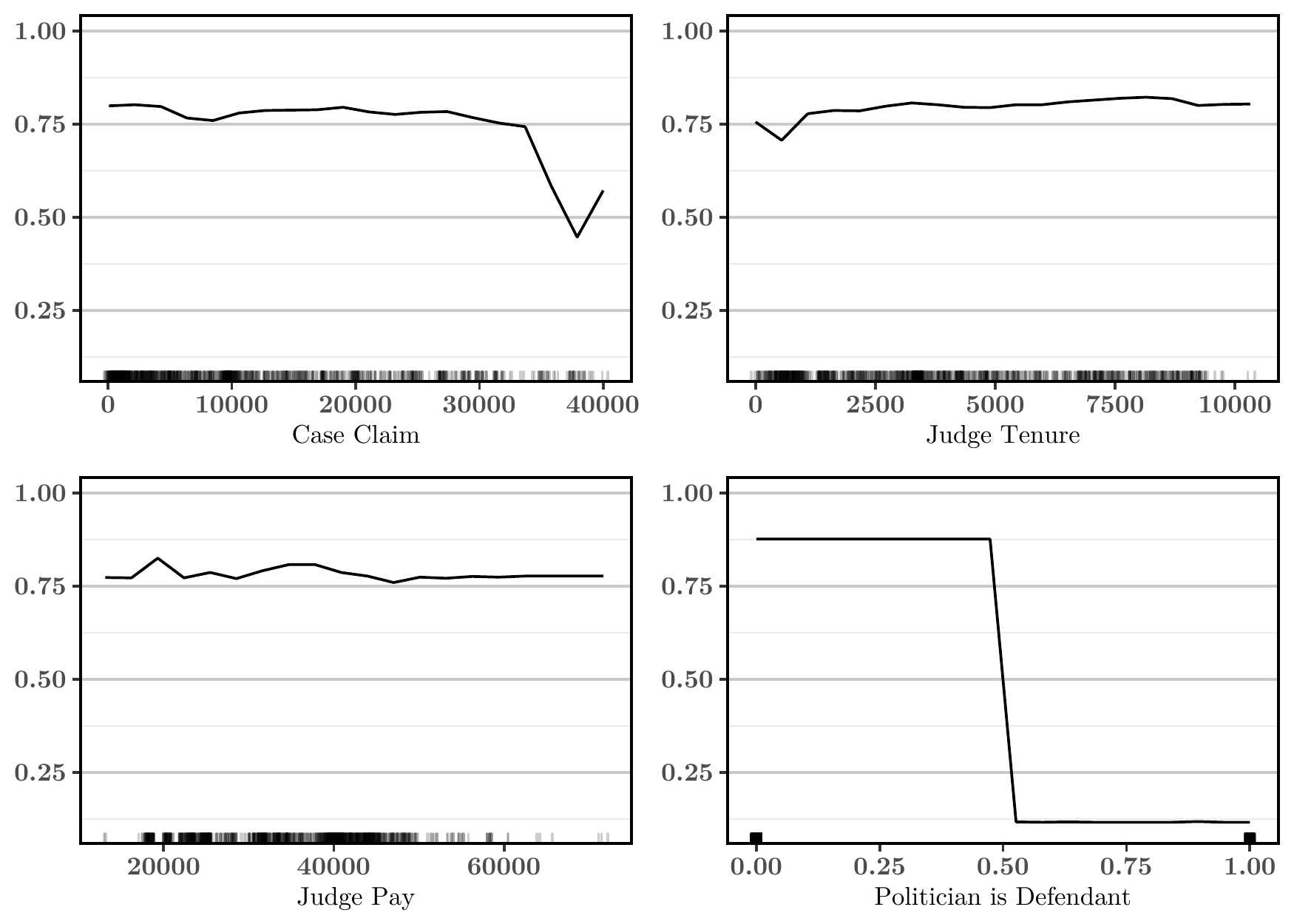}
\end{figure}

In addition to Gini, we also present evidence from partial dependency plots (PDP) in figure \ref{fig:pdp}. These graphs depict the distribution of marginal effects of any variable in a Random Forest regression, allowing us to identify the direction of the effect of each factor on court outcome. PDPs recover the marginal effect of an explanatory variable $x_s$ on the outcome of the model by integrating $x$ over the range of values for all other covariates $X_{-s}$. In the case of a linear model, for example, the PDPs would display a single linear effect. Figure \ref{fig:pdp} displays the PDPs of the four most important variables detected in the random forest model. The \emph{y}-axis displays how the probabilities of politicians winning their SCT cases change with marginal changes on each of the four variables. The shaded area under the \emph{x}-axis shows the concentration of observations in the hold-out data at each value, and larger blocks indicate more observations at a particular level of the explanatory variable.

We analyze marginal effects by looking at the trend for each variable in figure \ref{fig:pdp}. Pro-politician rulings have a downward trend over the range of case claim values. SCT cases whose claimants are asking for higher compensations are more exposed to unfavorable court rulings. Judge tenure is slightly upward-slopping, and serves as an indication that more experienced judges tend to favor politicians in court. One potential explanation for this relationship is that judges with longer tenures have also been playing the politician litigation game for longer and have come to realize that cooperation, rather than punishment, might work best for their interactions with the other branches of government. This explanation is an interesting avenue for future research. The graphs are less informative for the effects of judge pay and defendant politicians. Marginal effects for judge pay are relatively stable over the range of wages, and defendant status unequivocally reduces the probability of winning a small claims case. This effect was expected, however, since there are only two possible values for this variable (0 or 1) and small claims cases are noticeably decided in favor of claimants.

\section{Conclusion}\label{sec:conclusion_paper2}

This paper investigates whether there is differential treatment of politicians in small claims courts in the state of S\~ao Paulo, Brazil. To our knowledge, it is the first paper to produce clear evidence against judicial independence in low-stakes settings. State judges are arguably using low-salience rulings to signal power to politicians, and perhaps to buy future cooperation if these politicians take up office. In particular, this behavior occurs in cases where politicians are the defendants in small claims cases, where they see a 3.3 percentage point lower win rate when compared to all other defendants.

Similar studies in the literature document a positive bias in court when politicians are defendant in corruption cases \citep{LambaisJudicialSubversionEvidence2018} or when they share the affiliation with the ruling party \citep{Sanchez-MartinezDismantlingInstitutionsCourt2017,Poblete-CazenaveCrimePunishmentPoliticians2019}. We supplement the literature by looking at the entire distribution of politicians, even those not holding office at the time of trial, and focusing on low-salience cases -- precisely where there should not exist any bias. Besides the evidence of judicial bias in cases involving politicians, this project also makes predictions of court outcomes and pinpoints the most critical factors driving judicial decisions: amount claimed in court, judge tenure, judge pay, defendant politicians, and politician vote share. Though we only focus on a small set of cases, those filed in S\~ao Paulo's special civil tribunals, the predictions should serve as a benchmark for the deviations in other judiciary systems. The majority of studies in the political economy literature posits that an independent judiciary is crucial for checking the power of the executive and the legislative, and for supporting economic development \citep[][and many others]{BalandChapter69Governance2010}. Thus, we hope this study provides valuable insights for policymakers in developing countries sharing similar institutional designs.

Future projects should adopt new strategies to identify the causal effect of holding office on court outcomes. We find a positive correlation for mayor candidates, but we are unable to partial out potential unobservable biases driving the positive effect. A mayor's ability, or the experience they have with state court systems, could simultaneously influence their office prospects and court outcomes. Thus, we only provide the magnitude of the association between holding office and pro-politician rulings but make no causal claim about this relationship. Projects identifying the causal effect or widening the scope to also include the effect for city council candidates would be great contributions to this literature.

\pagebreak

\pagebreak

\section*{Tables and Figures} \label{sec:tables_paper2}

\begin{table}[H]
\centering
\caption[table1]{Descriptive Statistics}\label{tab:sumstats_paper2}
\vspace{-10pt}

\centering
\scriptsize
\setlength{\tabcolsep}{-2pt}
\begin{tabular}{@{\extracolsep{10pt}}lccccc}
\\[-1.8ex]\hline
\hline \\[-1.8ex]
& \multicolumn{1}{c}{\emph{n}} & \multicolumn{1}{c}{Mean} & \multicolumn{1}{c}{St. Dev.} & \multicolumn{1}{c}{Min} & \multicolumn{1}{c}{Max} \\
\hline \\[-1.8ex]
\\[-1.8ex]
\emph{Case Level} & \multicolumn{5}{c}{} \\
\hspace{4pt} Case Duration (in days) & 5,224 & 361    & 433    & 1  & 5,416 \\
\hspace{4pt} Amount Claimed (in R\$) & 5,224 & 11,750 & 10,633 & 35 & 40,000 \\
\hspace{4pt} Pro-Politician Ruling   & 5,224 & .645   & .478   & 0  & 1 \\
\\[-1.8ex]
\emph{Judge Level} & \multicolumn{5}{c}{} \\
\hspace{4pt} Male             & 514 & .611   & .488   & 0      & 1 \\
\hspace{4pt} Tenure (in days) & 514 & 4,385  & 2,892  & 13     & 12,987 \\
\hspace{4pt} Wage (in R\$)    & 514 & 35,152 & 10,797 & 13,156 & 145,616 \\
\\[-1.8ex]
\emph{Candidate Level} & \multicolumn{5}{c}{} \\
\hspace{4pt} Age                            & 2,943 & 44     & 10.3    & 18 & 78 \\
\hspace{4pt} Male                           & 2,943 & .901   & .298    & 0  & 1 \\
\hspace{4pt} Political Experience           & 2,943 & .243   & .430    & 0  & 1 \\
\hspace{4pt} Elected to Office              & 2,943 & .402   & .500    & 0  & 1 \\
\hspace{4pt} Campaign Expenditures (in R\$) & 2,943 & 18,212 & 77,890 & 11 & 1,770,315 \\
\hline \\[-1.8ex]
\hline
\end{tabular}

\end{table}

\begin{table}[H]
\centering
\caption[table2]{Individual Identifier for Lawsuits in Brazil}\label{tab:numerocnj}
\vspace{-10pt}

\centering
\setlength{\tabcolsep}{-2pt}
\begin{tabular}{@{\extracolsep{10pt}}crlc}
\\[-1.8ex]\hline
\hline \\[-1.8ex]
\ldelim\{{6}{5cm}[$3002615$--$80.2013.8.26.0510$]
& 3002615: & case filing order \\
& 80:      & control sequence  \\
& 2013:    & filing year       \\
& 8:       & court system ID   \\
& 26:      & state ID          \\
& 0510:    & district ID       \\
\\[-1.8ex]\hline
\hline \\[-1.8ex]
\end{tabular}

\end{table}

\begin{table}[H]
\centering
\caption[table3]{Descriptive Statistics Across Politician and Non-Politician Sample}\label{tab:sumstas_random}
\vspace{-10pt}

\centering
\scriptsize
\setlength{\tabcolsep}{-2pt}
\begin{tabular}{@{\extracolsep{10pt}}lccccc}
\\[-1.8ex]\hline
\hline \\[-1.8ex]
\multicolumn{6}{c}{\emph{Panel A: Non-Politician Sample}} \\
\hline \\[-1.8ex]
& \multicolumn{1}{c}{\emph{n}} & \multicolumn{1}{c}{Mean} & \multicolumn{1}{c}{St. Dev.} & \multicolumn{1}{c}{Min} & \multicolumn{1}{c}{Max} \\
\hline \\[-1.8ex]
\\[-1.8ex]
\emph{Case Level} & \multicolumn{5}{c}{} \\
\hspace{4pt} Case Duration (in days) & 3,233 & 353   & 380  & 1 & 3,336 \\
\hspace{4pt} Amount Claimed (in R\$) & 3,233 & 9,565 & 9,631& 10 & 40,000 \\
\hspace{4pt} Claimant Win Rate       & 3,233 & .749 & .434 & 0 & 1 \\
\\[-1.8ex]
\emph{Judge Level} & \multicolumn{5}{c}{} \\
\hspace{4pt} Male             & 389 & .622   & .485   & 0      & 1 \\
\hspace{4pt} Tenure (in days) & 389 & 4,297  & 2,824  & 64     & 12,987 \\
\hspace{4pt} Wage (in R\$)    & 389 & 35,237 & 11,334 & 13,156 & 145,616 \\
\\[-1.8ex]
\\
\multicolumn{6}{c}{\emph{Panel B: Mean Difference Across Samples}} \\
\hline\\[-1.8ex]
& Mean & Mean & & & \\
& Politician & Non-politician & Mean & & \\
& Sample & Sample & Difference & \emph{t}-statistic & \emph{p}-value \\
& ($n =$ 5,224) & ($n =$ 3,233) & & & \\
\hline \\[-1.8ex]
\\[-1.8ex]
\emph{Case Level} & \multicolumn{5}{c}{} \\
\hspace{4pt} Case Duration (in days) & 360    & 353   & \mc{7.516}         & \mc{.837}   & .402 \\
\hspace{4pt} Amount Claimed (in R\$) & 11,749 & 9,565 & \hspace{-4pt}2,184 & \mc{9.737}  & .000 \\
\hspace{4pt} Claimant Win Rate       & .729   & .749  & \mc{-.019}         & \mc{-1.979} & .048 \\
\\[-1.8ex]
\\[-1.8ex]
\emph{Judge Level} & \multicolumn{5}{c}{} \\
\hspace{4pt} Male             & .611   & .622   & \mc{-.011}   & \mc{-.343} & .732 \\
\hspace{4pt} Tenure (in days) & 4,384  & 4,297  & \mc{87.055}  & \mc{.454}  & .650 \\
\hspace{4pt} Wage (in R\$)    & 35,151 & 35,237 & \mc{-86.074} & \mc{-.115} & .908 \\
\hline \\[-1.8ex]
\hline
\end{tabular}

\end{table}

\begin{table}[H]
\centering
\caption[table4]{Mean Difference of Outcomes at the Judge-Level}\label{tab:agg_outcomes}
\vspace{-10pt}

\centering
\scriptsize
\setlength{\tabcolsep}{-2pt}
\begin{tabular}{@{\extracolsep{10pt}}lccccc}
\\[-1.8ex]\hline
\hline \\[-1.8ex]
& Mean & Mean & & & \\
& Politician & Non-politician & Mean & & \\
& Sample & Sample & Difference & \emph{t}-statistic & \emph{p}-value \\
\hline \\[-1.8ex]
\multicolumn{1}{r}{Any Claimant $\times$ \hspace{1pt} Any Claimant} & .741 & .751 & $-$.010 & \hspace{3pt}$-$.616 & .538 \\
& {\tiny ($n = 514$)} & {\tiny ($n = 389$)} & & & \\\\[-1.8ex]
\multicolumn{1}{r}{Politician is Claimant $\times$ \hspace{1pt} Any Claimant}   & .729 & .751 & $-$.022 & $-$1.377 & .169 \\
& {\tiny ($n = 483$)} & {\tiny ($n = 389$)} & & & \\\\[-1.8ex]
\multicolumn{1}{r}{Any Defendant $\times$ Any Defendant} & .259 & .249 & \hspace{6pt}.010 & \hspace{10pt}.651 & .515 \\
& {\tiny ($n = 514$)} & {\tiny ($n = 389$)} & & & \\\\[-1.8ex]
\multicolumn{1}{r}{Politician is Defendant $\times$ Any Defendant}    & .216 & .249 & $-$.033 & $-$1.652 & .099 \\
& {\tiny ($n = 321$)} & {\tiny ($n = 389$)} & & & \\\\[-1.8ex]
\hline \\[-1.8ex]
\hline
\end{tabular}

\end{table}

\begin{table}[H]
\centering
\caption[table4]{Performance Measures For All Models}\label{tab:prediction}
\vspace{-10pt}

\centering
\scriptsize
\setlength{\tabcolsep}{-2pt}
\begin{tabular}{@{\extracolsep{10pt}}lcc}
\\[-1.8ex]\hline
\hline \\[-1.8ex]
\emph{Model}            &Accuracy & Kappa \\
\hline \\[-1.8ex]
\\[-1.8ex]
\hspace{2pt} 1. Random Forest        & 76.97\% & 19.21\% \\
\hspace{2pt} 2. Gradient Boosting    & 74.86\% & 15.92\% \\
\hspace{2pt} 3. Lasso                & 72.64\% & 12.49\% \\
\hspace{2pt} 4. Deep Neural Networks & 67.82\% & \hspace{4pt}5.03\% \\
\\[-1.8ex]
\hline \\[-1.8ex]
\hline
\end{tabular}

\end{table}

\pagebreak

\end{document}